\documentclass[conference,10pt]{IEEEtran}
\IEEEoverridecommandlockouts
\usepackage{cite}
\usepackage{amsmath,amssymb,amsfonts}
\usepackage{algorithmic}
\usepackage{graphicx}
\usepackage{textcomp}
\usepackage{xcolor}
\usepackage{pgfplots}
\usepackage{hyperref}

\usepackage{cleveref}
\usepackage{mathmacrosv2}
\usepackage{subcaption}
\pgfplotsset{compat=1.18}

\def\ac{\alpha_c}
\def\as{\alpha_s}

\def\herm{^\mathrm{H}}
\def\Pr{\mathtt{Pr}}

\def\lf{\left}
\def\ri{\right}

\def\oc{\omega_c}
\def\os{\omega_s}

\newcommand{\Span}[1]{\langle#1\rangle}
\newcommand{\ds}{d^{(s)}}
\def\gin{\gamma_{\rm min}}

\def\BibTeX{{\rm B\kern-.05em{\sc i\kern-.025em b}\kern-.08em
    T\kern-.1667em\lower.7ex\hbox{E}\kern-.125emX}}

\theoremstyle{plain}
\newtheorem{remark}{Remark}
\theoremstyle{plain}
\newtheorem{definition}{Definition}

\begin{document}

\title{Subspace-Coding for Bistatic ISAC: Taming Interference via Array Geometry

\thanks{\textsuperscript{*}
Equal contribution. This work was supported in part by DOE grant DE-SC0022165, NSF grant 2124929, CWC, and RCF grant 375331.
}
}

\author{\IEEEauthorblockN{Parthasarathi Khirwadkar\textsuperscript{*}}
\IEEEauthorblockA{{ECE Department} \\
{UC San Diego}}
\and
\IEEEauthorblockN{Mohamed Ghada-Waeel Hamdy\textsuperscript{*}}
\IEEEauthorblockA{{ECE Department} \\
{UC San Diego}}
\and
\IEEEauthorblockN{Robin Rajam\"{a}ki}
\IEEEauthorblockA{{Signal Processing Research Centre} \\
{Tampere University}}
\and
\IEEEauthorblockN{Piya Pal}
\IEEEauthorblockA{{ECE Department} \\
{UC San Diego}}
}

\maketitle

\begin{abstract}
    This paper introduces a subspace coding perspective to bistatic integrated sensing and communication (ISAC), highlighting the role of receiver array geometry in the coupling between communication decoding and sensing. A central insight is that the impact of the interference caused by the sensing channel on the line-of-sight communication path is controlled by the subspace distance between the sensing and communication subspaces, {\em which strongly depends on the array geometry}. By studying a commonly used successive decoding approach, we derive upper and lower bounds on the communication error probability in terms of an effective SINR, which is explicitly determined by this subspace distance. This reveals that array geometries designed to form well-separated sensing subspace codes can also improve communication reliability in bistatic ISAC. In particular, Bose--Chowla arrays,  shown to be near-optimal subspace codes for spatial sensing in our earlier works, can outperform uniform linear arrays in terms of symbol decoding errors. The performance gap is significant when the communication and sensing channels are strongly correlated, and it increases with the number of antennas. Numerical experiments confirm that subspace-coded array geometries provide more favorable sensing--communication tradeoffs than ULAs for iterative joint target localization and decoding (JTD) receiver. 
\end{abstract}


\section{Introduction}

\IEEEPARstart{I}{ntegrated} sensing and communication (ISAC) is a promising technology for next-generation wireless networks, enabling communication infrastructure to support both high-rate data connectivity and sensing of the physical environment \cite{Eldar2022jwa}. Such capability is central to 6G applications including UAV localization \cite{wu2021comprehensive}, vehicle-to-everything (V2X) communication \cite{liu2020radar}, autonomous driving\cite{ma2020joint}, and environment-aware beam management \cite{gonzalez2024integrated}. ISAC architectures are commonly classified as monostatic or bistatic, depending on whether the transmitter (Tx) and sensing receiver (Rx) are colocated or separated, respectively \cite{jiao2025information}. 
Although monostatic ISAC benefits from waveform knowledge and synchronization, it may require substantial full-duplex hardware modifications and self-interference mitigation at the base station \cite{barneto2021full}. 

{This motivates bistatic ISAC, where sensing is performed at an Rx separated from the Tx. Such architectures can reuse the distributed Tx/Rx infrastructure, communication signals, and hardware of existing wireless networks with limited modifications \cite{wu2024sensing,mishra2019toward}. Recent works have studied bistatic ISAC from both theoretical and algorithmic perspectives, highlighting sensing--communication tradeoffs and the role of joint sensing and decoding \cite{fodor2025trade-off-d6e,zhao2025joint-b71, liu2026bilinear-e85,shen2026fundamental-c8b,jiao2025information,Bloch2023,gupta2025data-aided-e4d}.} In particular, uplink bistatic ISAC has recently gained attention where the base station reuses the user-transmitted uplink pilot and data symbols for joint target localization and communication  \cite{zhang2017framework,zhao2025joint-b71,liu2026bilinear-e85,fodor2025trade-off-d6e}.  Recent single-RF-chain millimeter-wave prototypes further demonstrate the  practical viability of this architecture \cite{guo2026panoptic-f1d}.  Unlike downlink monostatic ISAC settings, where alignment between sensing and communication directions  helps reuse the transmit waveform \cite{lu2023performance}, the same spatial coupling can be detrimental in uplink bistatic ISAC.  Specifically, the target-reflected signal  can act as structured interference for the data detection \cite{fodor2025trade-off-d6e}. A plethora of recent works on \emph{monostatic} ISAC have 
investigated antenna selection and sparse arrays \cite{sankar2024sparse,rajamaki2024sparse,van2024receiver,wang2018dual,liu2024doa,rajamaki2026impossibility}, 
owing to their ability to enhance angular resolution and identifiability while using fewer spatial resources \cite{wang2016coarrays,amin2016sparse,Pal20234bj}. 
The role of sparse arrays in bistatic ISAC, however, remains less understood.
Related work has studied sparse arrays for multi-user MIMO and integrated localization and communication systems \cite{wang2026enhancing,min2025integrated,li2025sparse}. 
In these settings, sparse arrays can reduce inter-user interference compared to standard ULAs by narrowing the mainlobe of the array beampattern, thereby reducing the correlation between steering vectors associated with different users\cite{wang2026enhancing,min2025integrated}. In particular, \cite{min2025integrated} compared various sparse array geometries using beampattern-based descriptors, such as beamwidth and sidelobe level. 
However, these do not directly translate into reliability guarantees for decoding or sensing in bistatic ISAC, thus leaving the role of array geometry obscure.

In this paper, we make this connection explicit by analyzing the probability of decoding error of a simple ``decode-then-sense'' receiver. We show that the decoding error probability is governed by the receiver array geometry through the subspace distance between the line-of-sight communication steering vector and the target-reflected steering vector. Building on the subspace coding perspective of our prior work \cite{mahdavifar2024isit}, we show that Bose--Chowla arrays, which form near-optimal sensing subspace codes, also improve communication reliability. In particular, even for closely spaced user and target DoAs, the decoding error probability decays exponentially with the number of sensors, whereas in case of the ULA, this 
remains bounded away from zero.

\section{Problem Formulation and Background 
}

\label{Sec_Problem_Formulation}

We consider an  uplink bistatic ISAC setting between a base station (BS) with $M$ antennas, and a single-antenna user equipment (UE) in the presence of a single sensing target (as in, e.g., \cite{liu2021cramer,Hua,Caire-fundamental}). The UE transmits a block of \(T\)  communication symbols \(\x=[x_1,\ldots,x_T]\in\CC^{1\times T}\) from a fixed codebook $\cX$, where \(\|\x\|_2^2=E_x,\,\forall\x\in\cX\). The BS receives a superposition of signals from the line-of-sight (LOS) communication path and a reflection from the sensing target. The LOS signal impinges on the BS from direction-of-arrival (DOA) \(\omega_c \triangleq \sin(\theta_c)\), while the
DOA of the 
signal reflected from the target 
is 
\(\omega_s \triangleq \sin(\theta_s)\). The BS measurements can be collected into $ M\times T$ matrix 
\begin{align}
    \Y &= \left( \alpha_c\a_{\DD}\lf(\omega_c\ri) + \alpha_s\a_{\DD}\lf(\omega_s\ri) \right)\x + \N,\label{eq:matrix_model}
\end{align}
where \(\a_{\DD}(\omega)\in\CC^{M\times 1}\) is the array steering vector in direction $\omega =\sin(\theta)$ with entries \([\a_{\DD}(\omega)]_m=e^{j\frac{2\pi}{\lambda} d_m\omega},\ m\in [M]\), \(\DD=\frac{\lambda}{2}\{d_1,d_2,\dots,d_M\}\) are the sensor locations of the (linear) array at the BS, and $\lambda$ is the carrier wavelength. Furthermore, \(\ac\) and \(\as\) are the complex gains of the LOS path and target-reflected path, respectively, and \([\N]_{:,t}\overset{ iid}{\sim}\CN(\zero,\sigma^2\I_M),\forall t\).

We assume that the LOS channel \(\alpha_c\a_{\DD}(\omega_c)\) is known at BS from a prior pilot-based channel estimation phase (as in \cite{liu2024doa,lu2023performance,Eldar2022jwa}), whereas \(\omega_s,\as\) and the transmitted codeword $\x\in \cX$ are unknown to the BS. We further assume that 
\(\alpha_s\sim\CN(0,\sigma^2_s)\) and $\x$ is selected uniformly at random from $\cX$ \cite{liu2026bilinear-e85}. A central goal of bistatic ISAC is to decode the communication symbol $\x$, while simultaneously estimating the target direction $\omega_s$. These objectives are inherently conflicting, since the unknown sensing signal acts as interference to symbol decoding. {\em The goal of this paper is to investigate whether the array geometry can be optimized to balance this tradeoff and provably improve ISAC performance.} We begin by reviewing our recent work on optimal (subspace) codes for sensing.
\vspace{-12pt}
\subsection{Review of subspace coding perspective of spatial sensing }
\vspace{-2pt}
{ In \cite{mahdavifar2024isit}, we made the first connection between sparse arrays and optimum one-dimensional subspace codes, a relationship that will prove highly relevant to bistatic ISAC.} Notice that for the pure sensing problem ($\alpha_c=0$), since $\alpha_s$ is unknown, the information about target DOA $\omega_s=\sin(\theta_s)$ is (unambiguously) encoded in the column space of the received measurement $\Y$.

 Our earlier work \cite{mahdavifar2024isit} showed that the receiver array geometry controls the ability to distinguish the true target subspace $\Span{\a_{\DD}(\omega_s)}\triangleq \mathrm{span}\{\a_{\DD}(\omega_s)\}$ from the subspaces of other candidate DoAs, $\Span{\a_{\DD}(\omega')}$, $\omega'\neq \omega_s$. 
 
 Assuming that the set of candidate DoAs lie on a uniform grid $\Omega\subset[-1,1)$ of $ |\Omega| = N$ spatial frequencies,  the set of candidate subspaces $\cC_{\DD} = \{\Span{\a_{\DD}(\omega)}:\omega\in\Omega\}$ can be considered a codebook of DOA-indexed subspaces. The reliability of identifying the correct DoAs then depends on the separation among the codewords in $\cC_{\DD}$. This separation is quantified by the ``subspace (quasi-)distance" between any two subspace codewords $\Span{\a_{\DD}(\omega)}$ and $ \Span{\a_{\DD}(\omega')}$ defined as follows \cite{Mahdavifar2022,mahdavifar2024isit}
\begin{align}
\ds\left(\Span{\a_{\DD}(\omega)}, \Span{\a_{\DD}(\omega')}\right)
\triangleq
1-\frac{1}{M^2}
\left|\a_{\DD}(\omega)\herm\a_{\DD}(\omega')\right|^2 .
\label{Eq_subspace_dist}
\end{align}
For brevity, we will use the notation $\ds_{\DD}(\omega,\omega') := \ds(\Span{\a_{\DD}(\omega)},\Span{\a_{\DD}(\omega')})$. It is shown in \cite{mahdavifar2024isit} that the minimum pairwise subspace distance determines the probability of error in estimating the DOA. This subspace-coding viewpoint motivates the use of sparse array geometries, which maximize the following minimum subspace distance 
\begin{align}
    \ds_{\min}(\cC_{\DD}) \triangleq \min_{(\omega,\omega')\in\Omega,\ \omega\neq\omega'} \ds_{\DD}\left(\omega,\omega'\right).
    \label{eq:min_subspace_dist}
\end{align}

It is shown in \cite{mahdavifar2024isit} that the following sparse array based on Bose--Chowla construction achieves near optimal minimum subspace distance over a uniform grid $\Omega$ of size $N\triangleq M^2-1$.

\begin{definition}[Bose--Chowla array]
Let \(M=p^k\) be a power of a prime \(p\), and let \(\beta\) denote a primitive element of the finite field \(\FF_{M^2}\). The sensor locations of the BCA are given by 
\begin{align}
    \BB =\frac{\lambda}{2}\{i\in[M^2-2]: \beta^i-\beta \in \FF_M \}.
    \label{Eq_BCA}
\end{align}    
\end{definition}
The minimum subspace distance of the Bose--Chowla array (BCA) satisfies
\vspace{-13pt}
\begin{align}
    \ds_{\min}(\cC_{\BB}) > 1 - \frac{2}{M},
    \label{Eq_BCA_min_dist}
\end{align}
which is within constant factor of the Welch bound \cite{welch1974lower} of $\approx 1 - \frac{1}{M}$.
{While \cite{mahdavifar2024isit} studied subspace codes in a sensing-only setting, we show that subspace distance directly influences the detection error rate of $\x$ in ISAC. As a result, near-optimal constructions such as Bose–Chowla codes can significantly improve performance, further highlighting the importance of array geometry.}

\vspace{-5pt}
\section{Array Geometry and Subspace Coding in Bistatic ISAC}
\label{sec_results}

In order to demonstrate the role of array geometry in ISAC, we focus on a commonly used joint sensing and communication strategy, namely decode-then-sense (DTS), which first decodes the communication symbols and then uses the decoded data to aid target sensing. This has been studied as the prototypical strategy for bistatic ISAC \cite{shen2026fundamental-c8b,Bloch2023} which forms the basis for other iterative joint sensing and communication decoding \cite{zhao2025joint-b71,liu2026bilinear-e85,gupta2025data-aided-e4d}. Indeed, DTS is a form of \emph{interference cancellation}, which is a widely employed technique in both ISAC \cite{Eldar2022jwa} and multiuser communications \cite{tse2005fundamentals}.

The DTS approach under consideration first estimates the communication codeword by solving the following problem

\begin{align}\label{eq:DTS_equivalent_decoder}
    \hat{\x} = 
    \argmin_{\widetilde{\x}\in \cX} \|\v - \widetilde{\x}\|_2^2,  
\end{align}
where $\v$ is the output of beamforming in the direction of $\omega_c$.

\begin{align}\label{eq:DTS_beamformer_output}
    \v \triangleq \frac{\lf(\alpha_c\a_{\DD}\lf(\omega_c\ri)\ri)\herm\Y}{\|\alpha_c \a_{\DD}(\omega_c)\|_2^2}  = \frac{\a_{\DD}(\omega_c)\herm\Y}{M\alpha_c}.
\end{align}
After decoding the communication codeword, the receiver subtracts the reconstructed LOS component from the received measurements and uses the residual $\widetilde{\Y} :=
       \Y - \alpha_c\a_{\DD}(\omega_c)\hat{\x}$ for sensing the target DOA. The target DOA is then estimated by solving the following problem 
\begin{align}
    \hat{\omega}_s
    =
    \argmin_{\omega\in\Omega}
    \min_{\beta\in\CC}
    \left\|
        \widetilde{\Y}
        -
        \beta \a_{\DD}(\omega)\hat{\x}
    \right\|_F^2 .
    \label{eq_LS}
\end{align}

Substituting the estimate of $\beta$ in terms of $\omega$ into \eqref{eq_LS} yields

\begin{align}
    \hat{\omega}_s = \argmax_{\omega\in\Omega}\left| \a_\DD\herm(\omega)\widetilde{\Y}\hat{\x}\herm \right|^2.
    \label{Eq_sensing}
\end{align}

Assuming $\omega_c$ is known, we denote the average probability of erroneously decoding a communications codeword $\x$ given \(\omega_s\) as \(
    P_e(\omega_s) \triangleq \Pr(\hat{\x} \neq \x; \omega_s).
    \) The following lemma relates the probability of decoding error with the subspace distance between $\Span{\a_{\DD}(\omega_c)}$ and $\Span{\a_{\DD}(\omega_s)}$, without restricting \(\omega_c\) to an angular grid.

\begin{lemma}[Bounds on DTS decoding error]
    \label{lem:Pe_comm_bounds}
    For each \(\x,\x'\in\cX\), \(\x'\neq\x\), let \(\d=\x'-\x\), and define the pairwise effective signal-to-interference-plus-noise ratio (SINR) as
    \begin{align}
        \!\!\!\!\gamma_{\x,\x'}(\DD)
        \!\triangleq\!
        \frac{
        M|\alpha_c|^2\|\d\|_2^4
        }{
        2\!\left(
        \sigma^2\|\d\|_2^2
        \!+\!\!
        M\sigma_s^2
        \bigl(1\!-\ds_\DD(\omega_c,\omega_s)\bigr)
        |\d\x\herm|^2
        \right)
        }.
        \label{eq:gamma_eff_pairwise}
    \end{align}
    Define \(
        \gamma_{\rm eff}(\DD)
        \triangleq
        \min_{\x,\x'\in\cX,\ \x'\neq\x}
        \gamma_{\x,\x'}(\DD).
    \) Let \(Q(x)=\frac{1}{\sqrt{2\pi}}\int_x^\infty e^{-u^2/2}\,du\). Then, the decoding error probability \(P_e(\omega_s)\) satisfies
    \begin{align}
        \!\!\!\frac{1}{|\cX|}Q\lf(\!\sqrt{\gamma_{\rm eff}(\DD)}\ri)\!\leq\!P_e(\omega_s)
        \!\leq\!\lf(|\cX|-1\ri)
        Q\!\lf(\!\sqrt{\gamma_{\rm eff}(\DD)}\ri),
        \label{eq:Pe_comm_bounds}
    \end{align}
     and has an exponential decay rate with increasing $\gamma_{\rm eff}(\DD)$ given by
    \(
        -\ln P_e(\omega_s) = \frac{1}{2}\gamma_{\rm eff}(\DD) + \Theta\left(\ln \gamma_{\rm eff}(\DD)\right).
    \)
\end{lemma}

We give a brief sketch of the proof here. Assume the correct codeword is \(\x\). Substituting in \cref{eq:DTS_beamformer_output} gives
\begin{align}
    \v = (1+\kappa)\x+ \bm{\eta}, \label{eq:apdx_dts_beamformer_output}
\end{align}
where \(\kappa \triangleq \frac{\alpha_s}{M\alpha_c} \bm{a}_\DD\herm(\omega_c)\bm{a}_\DD(\omega_s)\), and \( \bm{\eta} \triangleq \frac{1}{M\alpha_c}\bm{a}_\DD\herm(\omega_c)\N\sim\mathcal{CN}(0,\sigma_\eta^2\I_T)\) with \(\sigma_\eta^2 \triangleq \frac{\sigma^2}{M|\alpha_c|^2}\). The bounds in   \cref{eq:Pe_comm_bounds} are obtained by characterizing the probability of pairwise error event \(\mathtt E(\x,\bm x')\triangleq \left\{ \|\bm{v}-\bm{x}'\|_2^2 <  \|\bm{v}-\bm{x}\|_2^2 \right\}\) for \(\x'\neq\x\). When this event occurs, the decoder prefers \(\bm{x}'\) over the true codeword \(\bm{x}\). Rewriting $\mathtt E(\x,\bm x')$ using \cref{eq:apdx_dts_beamformer_output} and isolating the noise term gives \vspace{-0.3cm} \begin{align*}
    \Pr(\mathtt E(\x,\x')\mid \x,\alpha_s;\omega_s)
    =
    Q\left(
    \frac{\|\bm{d}\|_2^2-2\mathrm{Re}\{\kappa\bm{d}\herm\x\}}{\sqrt{2\sigma_\eta^2\|\bm{d}\|_2^2}}
    \right).
\end{align*}
Denote $\rho_{\x,\x'} \triangleq \mathrm{Re}\{\kappa\bm{d}\herm\x\}$. Since \(\alpha_s\sim\mathcal{CN}(0,\sigma_s^2)\), we have $\rho_{\x,\x'}|\x \sim \cN(0,\sigma_\rho^2)$ where 
\begin{align*}
     \sigma_\rho^2
    =
    \frac{\sigma_s^2}{2|\alpha_c|^2}
    \left(1-\ds_\DD(\omega_c,\omega_s)\right)
    |\bm{d}\herm\x|^2.
\end{align*}
To evaluate
    \[
    \Pr(\mathtt E(\x,\x')\mid \x;\omega_s)
    =
    \EE_{\alpha_s}
    \left[
    \Pr(\mathtt E(\x,\x')\mid \x,\alpha_s;\omega_s)
    \right],
    \]
we use the following identity
\begin{align*}
    &\EE_{X\sim \cN(0,\sigma_{\rho}^2)}\!
    \left[
    Q\left(\frac{A-2X}{B}\right)
    \right]
    \!=
    \!\EE_{X}\!\left[\Pr\left(Z > \frac{A-2X}{B} \Big| X\right) \right]\\
    &\quad =\Pr\left(BZ+2X>A\right)  =
    Q\left(\!
    A/\sqrt{B^2+4\sigma_\rho^2}
    \right),
\end{align*}
where \(Z\sim\mathcal{N}(0,1)\) is independent of \(X\). Substituting $X \triangleq \rho_{\x,\x'}$, \(A\triangleq \|\bm{d}\|_2^2\), and \(B\triangleq \sqrt{2\sigma_\eta^2\|\bm{d}\|_2^2}\) in the above identity gives the following
\vspace{-7pt}
\begin{align*}
    \Pr(\mathtt E(\x,\x')\mid\x; \omega_s)
    =
    Q\left(
    \sqrt{\gamma_{\x,\x'}(\DD)}
    \right),
\end{align*}
where \(\gamma_{\x,\x'}(\DD)\) is defined as in \cref{eq:gamma_eff_pairwise}. We obtain the upper bound in \cref{eq:Pe_comm_bounds} using union bound. Since \(Q(\cdot)\) is decreasing and $\gamma_{\x,\x'}(\DD)\geq \gamma_{\rm eff}(\DD)$ for all \(\x'\neq\x\), it follows that
\begin{align}
P_e(\omega_s) &=
    \frac{1}{|\cX|}
    \sum_{\x\in\cX}
    \Pr(\hat{\x}\neq\x\mid \x;\omega_s) \label{eq:avg_Pe_comm}\\
    &\leq \frac{1}{|\cX|} \sum_{\x\in \cX}\sum_{\x' \in \cX, \x'\neq \x} \Pr(\mathtt E(\x,\x')\mid\x; \omega_s) \nonumber \\
    &\leq (|\cX|-1)
    Q\left(
    \sqrt{\gamma_{\rm eff}(\DD)}
    \right). \nonumber
\end{align}
For the lower bound in \cref{eq:Pe_comm_bounds}, let \((\x_\star,\x'_\star)\) be a pair of distinct codewords attaining \(\gamma_{\rm eff}(\DD)\). We can lower bound the LHS of \cref{eq:avg_Pe_comm} as follows 
\begin{align*}
    \begin{split}
    P_e(\omega_s)
    &\geq\frac{1}{|\cX|}
    \Pr(\hat{\x}\neq\x_\star\mid \x_\star;\omega_s)
    \geq
    \frac{1}{|\cX|}Q\left(
    \sqrt{\gamma_{\rm eff}(\DD)}
    \right).
    \end{split}
\end{align*}
The exponential rate stated in \Cref{lem:Pe_comm_bounds} follows from the inequalities
\(Q(u)\leq \frac{1}{\sqrt{2\pi} u} e^{-u^2/2}\) and
\(Q(u)\geq \frac{1}{\sqrt{2\pi}}\frac{u}{1+u^2}e^{-u^2/2}\) for \(u\geq0\) \cite{gordon1941values}.

\begin{remark}

Lemma \ref{lem:Pe_comm_bounds} shows that the error exponent of \(P_e(\omega_s)\) is determined by 
the subspace distance between $\Span{\a_{\DD}(\omega_c)}$ and $\Span{\a_{\DD}(\omega_s)}$. This highlights the role of array geometry in improving the error exponent by increasing the subspace distance between communication and sensing subspaces.
\end{remark}

To elucidate the role of array geometry and derive explicit expressions for \(P_e(\omega_s)\), we will compare the performance of BCA and ULA in two regimes: (i) closely aligned sensing and communication channels, and (ii) spatially separated UE and sensing targets.

\vspace{-5pt}
\subsection{Closely aligned Communication and Sensing channels}

Let \(\Delta \triangleq |\omega_c-\omega_s|\) denote the angular separation between communication and sensing channels. We consider the regime where  \( \Delta = \Theta(\frac{1}{M^2}) \). This is a challenging scenario, where the sensing channel is strongly correlated with the communication channel, causing strong sensing-induced interference. 
\footnote{Such close alignment is also of particular interest in ISAC, as the benefits of fully integrating communication and sensing functionalities are generally expected to be greatest in this regime.} The following theorem establishes the superiority of BCA (near-optimal subspace codes) to ULA in this scenario.
\begin{theorem}
    \label{thm:Pe_vs_M_rate_bounds}
    Let \(\Delta=\frac{2}{M^2-1}\). Define
    \begin{align}
        \gin
        \triangleq
        \min_{\x,\x'\in\cX,\ \x'\neq\x}
        \frac{
        |\alpha_c|^2\|\d\|_2^4
        }{
        4\left(
        \sigma^2\|\d\|_2^2
        +
        2\sigma_s^2|\d\x\herm|^2
        \right)
        }.
        \label{eq:codebook_exp_constant}
    \end{align}
    Then, for the BCA ($\DD = \BB$ as given in \eqref{Eq_BCA})
    \vspace{-0.15cm}
    \begin{align}
        \liminf_{M\to \infty}
        -\frac{\ln P_{e,\BB}(\omega_s)}{M}
        \geq
        \gin.
        \label{eq:Pe_BCA_exp_rate}
    \end{align}
    On the other hand, for the ULA ($\DD=\UU \triangleq \frac{\lambda}{2}\{0,1,\dots,M-1\}$)
    \vspace{-0.45cm}
    \begin{align}
        \lim_{M\to \infty}
        -\frac{\ln P_{e,\UU}(\omega_s)}{M}
        =
        0 .
        \label{eq:Pe_ULA_exp_rate}
    \end{align}
\end{theorem}

Here we give a sketch of the proof for \Cref{thm:Pe_vs_M_rate_bounds}. Using \Cref{lem:Pe_comm_bounds}, since \(|\cX|\) is fixed, we have
\begin{align}
    -\frac{\ln P_{e,\DD}(\omega_s)}{M}
    \geq
    \frac{\gamma_{\rm eff}(\DD)}{2M}
    +
    \frac{\ln\left(\gamma_{\rm eff}(\DD)\right)}{2M}
    +
    O\left(\frac{1}{M}\right).
    \label{eq:thm1_rate_lower_bound}
\end{align}
For the BCA, using \eqref{Eq_BCA_min_dist}, we have  \(
    M\left(1-\ds_{\BB}(\omega_c,\omega_s)\right)<2 .
    \) Hence, for any pair \(\x,\x'\in\cX\), \(\x'\neq\x\),
\begin{align*}
    \begin{split}
            \frac{\gamma_{\x,\x'}(\BB)}{M}
    &=
    \frac{
    |\alpha_c|^2\|\d\|_2^4
    }{
    2\left(
    \sigma^2\|\d\|_2^2
    +
    M\sigma_s^2
    \bigl(1-\ds_{\BB}(\omega_c,\omega_s)\bigr)
    |\d\x\herm|^2
    \right)
    } \\
    &>
    \frac{
    |\alpha_c|^2\|\d\|_2^4
    }{
    2\left(
    \sigma^2\|\d\|_2^2
    +
    2\sigma_s^2|\d\x\herm|^2
    \right)
    } .
    \end{split}
\end{align*}
\vspace{-0.1cm}
Therefore,
\begin{align}
    \frac{\gamma_{\rm eff}(\BB)}{2M}
    >
    \min_{\substack{\x,\x'\in\cX\\\x'\neq\x}}
    \frac{
    |\alpha_c|^2\|\d\|_2^4
    }{
    4\left(
    \sigma^2\|\d\|_2^2
    +
    2\sigma_s^2|\d\x\herm|^2
    \right)
    }
    =
    \gin.
    \label{eq_snr_bca_limit}
\end{align}
Thus, taking the limit infimum on both sides of \eqref{eq:thm1_rate_lower_bound} after plugging in \eqref{eq_snr_bca_limit} gives \eqref{eq:Pe_BCA_exp_rate}.

Using the lower bound \(Q(x)\geq \frac{1}{\sqrt{2\pi}}\frac{x}{x^2+1}e^{-x^2/2}\)~\cite{gordon1941values}
along with \Cref{lem:Pe_comm_bounds}, we have
\begin{align}
    \!\!\!-\frac{\ln P_{e,\DD}(\omega_s)}{M}
    \!\leq
    \!\frac{\gamma_{\rm eff}(\DD)}{2M}
    \!\!+
    \!\frac{1}{M}
    \ln\!\!\left(\!
        \frac{\gamma_{\rm eff}(\DD)+1}
        {\sqrt{\gamma_{\rm eff}(\DD)}}
    \!\right)
    \!\!+\!
    O\!\!\left(\!\frac{1}{M}\!\right)\!.
    \label{eq:thm1_rate_upper_bound}
\end{align}
For the ULA, since \(\ds_{\UU}(\omega_s,\omega_c)\leq 1-\frac{4}{\pi^2}\)~\cite[eq.~(25)]{mahdavifar2024isit}, we obtain
\vspace{-8pt}
\begin{align}
    \begin{split}
    \gamma_{\rm eff}(\UU)
    \!\leq\!\min_{\substack{\x,\x'\in\cX\\\x'\neq\x}}
    \frac{
    M|\alpha_c|^2\|\d\|_2^4
    }{
    2\left(
    \sigma^2\|\d\|_2^2
    \!+\!
    M\sigma_s^2
    \frac{4}{\pi^2}
    |\d\x\herm|^2
    \right)
    }\!=\!O(1).
    \end{split}
    \label{eq_snr_limit_ula}
\end{align}
Hence, the right-hand side of \cref{eq:thm1_rate_upper_bound} is \(O(1/M)\) for the ULA. Since \(-\frac{\ln P_{e,\UU}(\omega_s)}{M}\geq 0\), the limit in \cref{eq:Pe_ULA_exp_rate} exists and equals~\(0\).

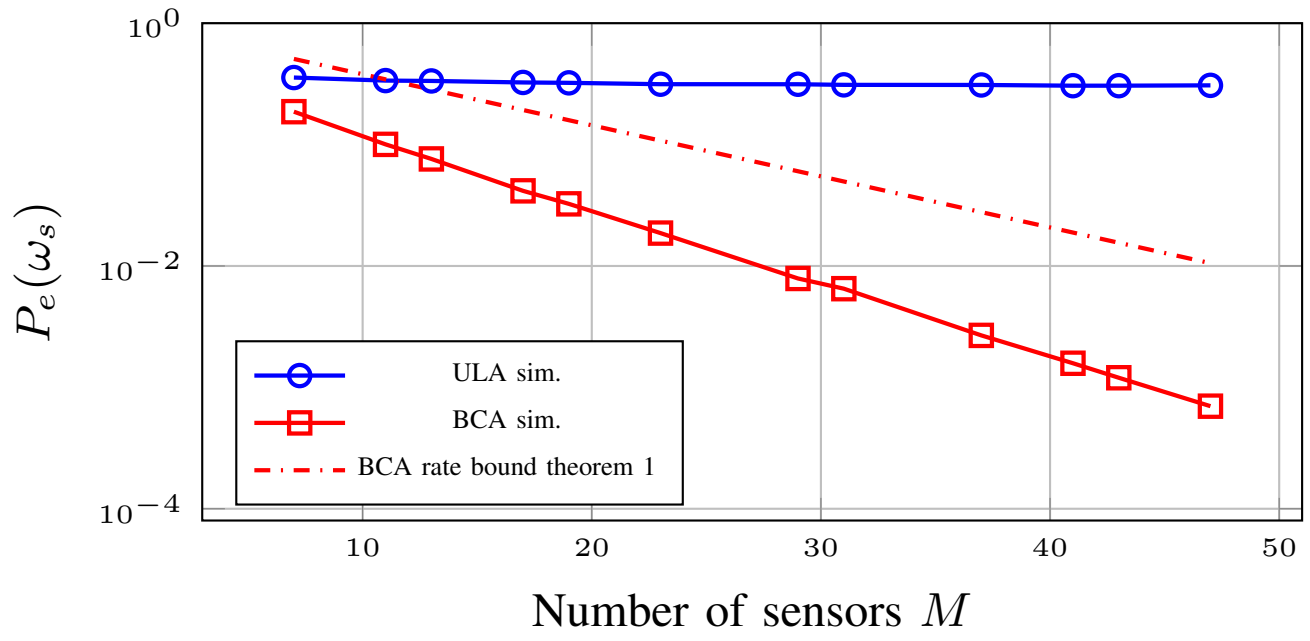
\begin{figure}[t]
    \centering
    \pgfplotstableread[col sep=comma]{data/dts-theorem1-pe-vs-m-qpsk1-results.csv}\dtsPeMTable
    \pgfplotstablegetelem{0}{SNR_dB}\of\dtsPeMTable
    \edef\dtsPeMSNR{\pgfplotsretval}
    \pgfplotstablegetelem{0}{Sigma2}\of\dtsPeMTable
    \edef\dtsPeMSigma{\pgfplotsretval}
    \pgfplotstablegetelem{0}{Theta_c_deg}\of\dtsPeMTable
    \edef\dtsPeMThetaC{\pgfplotsretval}
    \begin{tikzpicture}
        \begin{semilogyaxis}[
            width=\columnwidth,
            height=0.55\columnwidth,
            grid=both,
            xlabel={Number of sensors \(M\)},
            ylabel={$P_e(\omega_s)$},
            ymin=8e-5,
            ymax=1,
            unbounded coords=discard,
            tick label style={font=\tiny},
            label style={font=\small},
            legend style={font=\tiny, at={(0.03,0.03)}, anchor=south west},
        ]
            \addplot+[blue, mark=o, thick] table[col sep=comma, x=M,
                y=ULA_PeSim]
                {data/dts-theorem1-pe-vs-m-qpsk1-results.csv};
            \addlegendentry{ULA sim.}
            \addplot+[red, mark=square, thick] table[col sep=comma, x=M,
                y=BoseChowla_PeSim]
                {data/dts-theorem1-pe-vs-m-qpsk1-results.csv};
            \addlegendentry{BCA sim.}
            \addplot+[red, mark=none, dashdotted, thick] table[col sep=comma, x=M,
                y=BoseChowla_Theorem1PeRateBound]
                {data/dts-theorem1-pe-vs-m-qpsk1-results.csv};
            \addlegendentry{BCA rate bound \cref{thm:Pe_vs_M_rate_bounds}}
        \end{semilogyaxis}
    \end{tikzpicture}
    \caption{DTS decoding error probability versus \(M\) under the close-DOA scaling \(\Delta=2/(M^2-1)\), with \(\theta_c=\dtsPeMThetaC^\circ\), $|\alpha_c|=1$, and \(\mathrm{SNR}=10\log_{10}(|\alpha_c|^2/\sigma^2)=\dtsPeMSNR\) dB. We consider $T=1$ with QPSK constellation.
    }
    \label{fig:dts-pe-vs-m}
    \vspace{-16pt}
\end{figure}

\begin{remark} 
    \Cref{thm:Pe_vs_M_rate_bounds}
    shows that when \(\Delta = O(\frac{1}{M^2})\), the probability of error of BCA decays exponentially in \(M\) with a positive rate. In other words, using \eqref{eq_snr_bca_limit} and \Cref{lem:Pe_comm_bounds}, 
    \vspace{-6pt}
    \begin{align}
        \begin{split}
            P_{e,\BB}(\omega_s)
            &\leq
            (|\mathcal X|-1)
            \,Q\!\left(\!
            \sqrt{
            2M\gin
            }
            \right) 
            \\&
            \leq
            \frac{|\mathcal X|-1}{2}
            \,\exp\!\left(
            -M\gin
            \right),
        \end{split}
        \label{eq_eq_limit_upper_bound}
    \end{align}
    where the second inequality follows from the Chernoff bound \(Q(x)\leq \frac{1}{2}e^{-x^2/2}\). On the other hand, \eqref{eq_snr_limit_ula} gives \(\gamma_{\rm eff}(\UU)=O(1)\). Therefore, using \Cref{lem:Pe_comm_bounds}, we have
    \vspace{-0.1cm}
    \begin{align}
        P_{e,\UU}(\omega_s)
        =
        \Omega(1).
    \end{align}
    \vspace{-0.1cm}
    Thus, the ULA error probability is bounded away from zero even as \(M\to\infty\). This happens because, for fixed \(\sigma_s^2\), when the DoAs get closer, the ULA combines the sensing direction coherently causing a strong interference that is not mitigated by increasing the number of antennas. This behavior is confirmed in \Cref{fig:dts-pe-vs-m}, where the empirical BCA error probability decays rapidly with \(M\), while the ULA curve remains nearly constant. Note that in practice, \(M\) does not have to be very large for this behavior to emerge. Thus, for closely spaced communication and sensing directions, BCAs utilize their subspace-code structure for improved communication reliability, while ULAs remain limited by their poor subspace distance  properties.
\end{remark}
\vspace{-3pt}

Although \Cref{thm:Pe_vs_M_rate_bounds} is derived for $\Delta = \frac{2}{M^2-1}$, it can be shown that the results holds more generally for $\Delta=\Theta(\frac{1}{M^2})$ for Bose--Chowla arrays.

\vspace{-5pt}
\subsection{Spatially Well-separated Comm and Sensing Channels} The regime \(\Delta =\Omega(1/\sqrt{M})\) is considered well-separated since  the UE and target DoAs are separated far beyond the mainlobes of both the ULA and BCA. \footnote{Technically $\Delta \gg \frac{1}{M}$ is already outside the mainlobes of ULA and BCA. However, in the regime $\Delta=\Theta(\frac{1}{M^\zeta}), \frac{1}{2} < \zeta \leq 1$, the performances of ULA and BCA fluctuate and it is difficult to compare their performances.} For a ULA, the subspace distance between directions \(\omega_c,\omega_s\) is given by \cite{mahdavifar2024isit}
\begin{align*}
 \ds_{\UU}(\omega_c,\omega_s) = 1-\frac{1}{M^2} \frac{\left|\sin(\pi M\Delta/2)\right|^2} {\left|\sin(\pi \Delta/2)\right|^2}.    
\end{align*}
Using \(\left|\sin(\pi M\Delta/2)\right|\leq 1\) and \(\sin(\pi\Delta/2)\geq \Delta\) for \(0<\Delta\leq 1\), we obtain $
    \ds_{\UU}(\omega_c,\omega_s) \geq 1-\frac{1}{M^2\Delta^2}. 
    $ In particular, for any separation satisfying \(\Delta \geq 1/\sqrt{2M}\), we have 
    $
    \ds_{\UU}(\omega_c,\omega_s) \geq 1-\frac{2}{M}. 
    $
Thus, in the well-separated regime, the ULA also attains a large subspace distance comparable to that of the BCA. 
Let
\(
    P_{e,{\rm diff}}(\omega_s) = P_{e,\BB}(\omega_s) - P_{e,\UU}(\omega_s).
\) If \(P_{e,\mathrm{diff}}(\omega_s)\leq0\), then
\(P_{e,\BB}(\omega_s)\leq P_{e,\UU}(\omega_s)\), thus the BCA performs at least as well as the ULA in terms of decoding error. If \(P_{e,\mathrm{diff}}(\omega_s)>0\), then the ULA has a communication advantage, but this advantage is bounded as
\(
    P_{e,\mathrm{diff}}(\omega_s)
    =
    P_{e,\BB}(\omega_s)-P_{e,\UU}(\omega_s)
    \leq
    P_{e,\BB}(\omega_s),
\)
since \(P_{e,\UU}(\omega_s)\geq0\). As established in \cite{mahdavifar2024isit}, the BCA continues to satisfy the subspace-distance guarantee \(\ds_{\BB}(\oc,\os)>1-\frac{2}{M}\) for the entire spatial frequency grid $\Omega\subset [-1,1)$. Therefore, the BCA upper bound in \eqref{eq_eq_limit_upper_bound} holds in the well-separated regime and we obtain

\vspace{-10pt}
\begin{align*}
    P_{e,{\rm diff}}(\omega_s)\le \frac{|\mathcal X|-1}{2} \exp\left( -M\gin \right).
\end{align*}

Thus, in this regime, even if the detection error probability of ULA were to become smaller than that of BCA, the gap between their error decreases exponentially in \(M\).
Therefore, optimum subspace codes such as BCA provides a large gain in the close-separation regime, while remaining competitive with the ULA in the far-separation regime. This behavior is corroborated by our numerical simulations, presented next.

Note that a BCA with \(M\) sensors has a physical aperture length of \(O(M^2)\), while that of the ULA is of \(O(M)\). However, in a mmWave setting, the wavelengths are very short such that the aperture length is not a limitation. On the other hand, the power consumption of ADCs sets a limitation on \(M\) for a digital array. It is noteworthy that despite being a sparse array, BCA does not suffer from ambiguity, precisely because it is an almost-optimal subspace code.

\vspace{-4pt}
\section{Simulations}
\label{Sec_Sim}

\begin{figure}[t]
    \centering
    \begin{subfigure}[t]{\columnwidth}
    \centering
    \begin{tikzpicture}
        \begin{semilogyaxis}[
            width=\linewidth,
            height=0.5\linewidth,
            grid=both,
            xlabel={SNR [dB]},
            ylabel={Average $P_e(\omega_s)$},
            ymin=1e-3,
            ymax=1,
            xmin=0,
            xmax=16,
            unbounded coords=discard,
            legend style={font=\tiny, at={(0.03,0.03)}, anchor=south west},
            tick label style={font=\tiny},
            label style={font=\small},
        ]
            \addplot+[mark=o, thick] table[col sep=comma, x=SNR_dB,
                y expr={max(\thisrow{ULA_ProbCommError},1e-4)}]
                {data/jtd-continuous-doa-qpsk1-no-pilot-array-geometry-results.csv};
            \addlegendentry{JTD\cite{zhao2025joint-b71} with ULA}
            \addplot+[mark=square, thick] table[col sep=comma, x=SNR_dB,
                y expr={max(\thisrow{Bose_Chowla_ProbCommError},1e-4)}]
                {data/jtd-continuous-doa-qpsk1-no-pilot-array-geometry-results.csv};
            \addlegendentry{JTD\cite{zhao2025joint-b71} with BCA}
        \end{semilogyaxis}
    \end{tikzpicture}
    \caption{ Communication vs. SNR}
    \label{fig:jtd-snr-results_comm}
    \end{subfigure}

    \begin{subfigure}[t]{\columnwidth}
    \centering
    \begin{tikzpicture}
        \begin{semilogyaxis}[
            width=\linewidth,
            height=0.5\linewidth,
            grid=both,
            xlabel={SNR [dB]},
            ylabel={\(\Pr(|\widehat{\theta}_s-\theta_s|>0.25^\circ)\)},
            ymin=1e-1,
            ymax=1,
            xmin=0,
            xmax=16,
            unbounded coords=discard,
            legend style={font=\tiny, at={(0.03,0.03)}, anchor=south west},
            tick label style={font=\tiny},
            label style={font=\small},
        ]
            \addplot+[mark=o, thick] table[col sep=comma, x=SNR_dB,
                y=ULA_ProbSensingError]
                {data/jtd-continuous-doa-qpsk1-no-pilot-array-geometry-results.csv};
            \addlegendentry{JTD\cite{zhao2025joint-b71} with ULA}
            \addplot+[mark=square, thick] table[col sep=comma, x=SNR_dB,
                y=Bose_Chowla_ProbSensingError]
                {data/jtd-continuous-doa-qpsk1-no-pilot-array-geometry-results.csv};
            \addlegendentry{JTD\cite{zhao2025joint-b71} with BCA}
        \end{semilogyaxis}
    \end{tikzpicture}
    \caption{DOA vs. SNR}
    \label{fig:jtd-snr-results_sensing}
    \end{subfigure}
    \caption{Communication and sensing error probabilities for JTD receiver \cite{zhao2025joint-b71} with varying SNR and averaged over target DOA $\theta_s\sim \mathrm{Uniform}([-60^\circ,60^\circ])$. }

    \label{fig:jtd-snr-average-theta_s}
    \vspace{-18pt}
\end{figure}
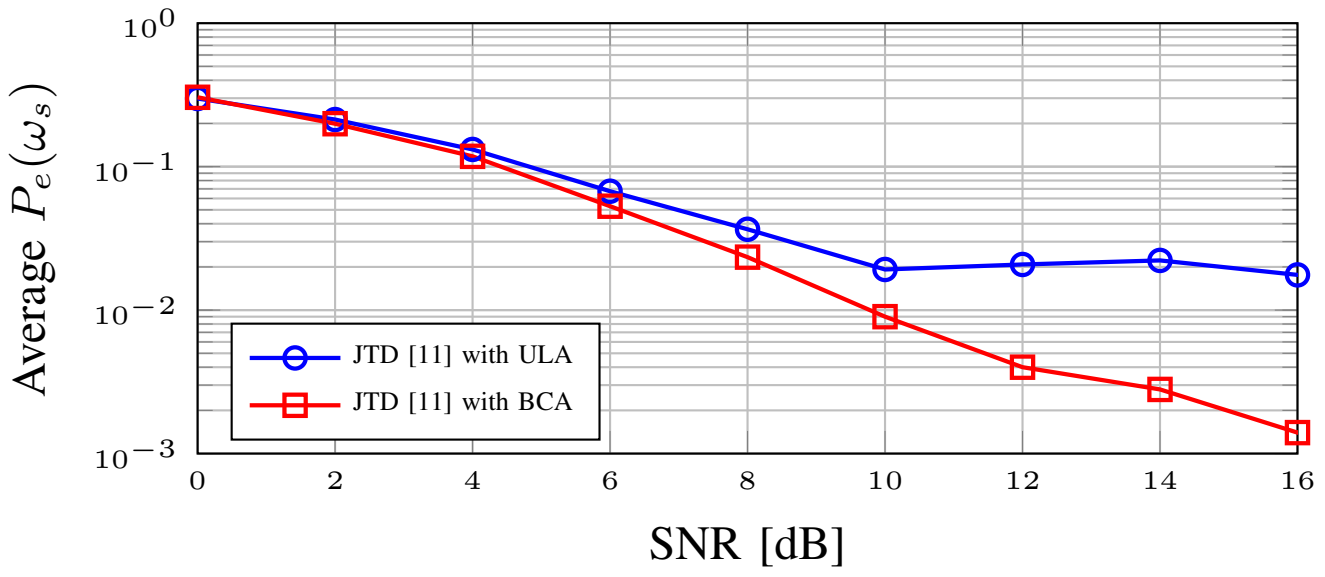
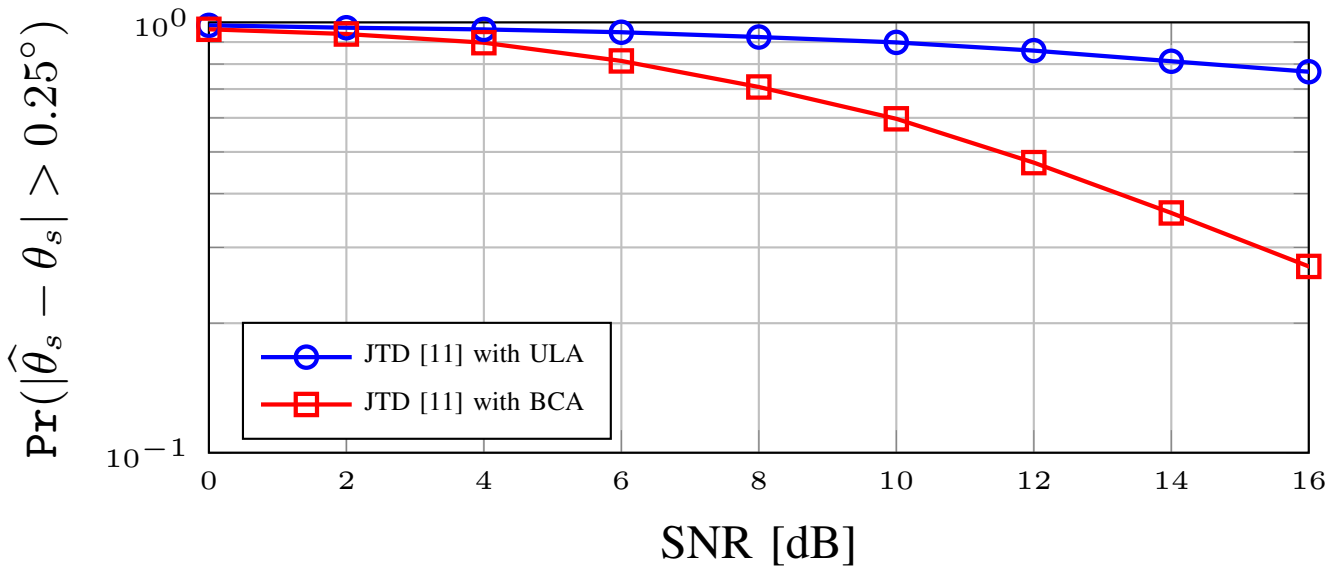

To test whether the insights obtained from the DTS analysis extend to more general joint receivers, we evaluate the joint target detection and decoding (JTD) method of~\cite{zhao2025joint-b71}. The JTD approach performs multiple iterations between decoding the unknown communication symbol using the current target estimates and then refining the target parameters using the detected data symbol. To match the model in  \eqref{eq:matrix_model}, 
we specialize JTD to DOA estimation with a single stationary target whose range is assumed to be known. Furthermore, the JTD receiver is given exact knowledge of \(\alpha_c\) and \(\theta_c\).

We compare JTD performance using ULA and BCA arrays with \(M=13\) sensors. Although the analysis in \cref{sec_results} is valid for any \(T\geq 1\), we use \(T=1\) communication symbol from a QPSK constellation for simplicity.
The first JTD iteration is initialized with the known LOS channel only; subsequent iterations estimate the target reflection from the decoded data symbol. Throughout the simulations, we set \(\alpha_c=e^{j\pi/4}\), draw \(\alpha_s\sim\mathcal{CN}(0,1)\), and use \(E_x=1\). The DTS SNR is defined as $\mathrm{SNR}\triangleq10\log_{10}\left(\frac{E_xM|\alpha_c|^2}{\sigma^2}\right).$

In \Cref{fig:jtd-snr-average-theta_s}, the UE direction is fixed at \(\theta_c=-10^\circ\), while the target DOA is drawn uniformly as \(\theta_s\sim\mathrm{Uniform}([-60^\circ,60^\circ])\). The DOA search grid has spacing \(0.5^\circ\), and a sensing error is declared if \(|\widehat{\theta}_s-\theta_s|>0.25^\circ\). As shown in \Cref{fig:jtd-snr-results_comm}, the ULA and BCA achieve comparable decoding error probabilities in this randomized DOA setting. In contrast, \Cref{fig:jtd-snr-results_sensing} shows that the BCA provides a substantial improvement in DOA sensing performance. This is consistent with the subspace-coding interpretation: the BCA improves target separability without degrading communication decoding.

\begin{figure}[t]
    \centering
    \begin{tikzpicture}
        \begin{semilogyaxis}[
            width=\columnwidth,
            height=0.5\columnwidth,
            grid=both,
            xlabel={\(\omega_s\)},
            ylabel={$P_e(\omega_s)$},
            ymin=1e-3,
            ymax=1,
            xmin=0,
            xmax=0.25,
            xtick={2/(13^2-1),1/13,1/sqrt(2*13)},
            xticklabels={\(\frac{2}{M^2-1}\),\(\frac{1}{M}\),\(\frac{1}{\sqrt{2M}}\)},
            unbounded coords=discard,
            legend style={font=\tiny, at={(0.97,0.97)}, anchor=north east},
            tick label style={font=\tiny},
            label style={font=\small},
        ]
            \addplot+[mark=o, thick] table[col sep=comma, x=Delta,
                y=ULA_ProbCommError]
                {data/jtd-delta-sweep-qpsk1-no-pilot-decoding-array-geometry-results.csv};
            \addlegendentry{JTD\cite{zhao2025joint-b71} with ULA}
            \addplot+[mark=square, thick] table[col sep=comma, x=Delta,
                y=Bose_Chowla_ProbCommError]
                {data/jtd-delta-sweep-qpsk1-no-pilot-decoding-array-geometry-results.csv};
            \addlegendentry{JTD\cite{zhao2025joint-b71} with BCA}
        \end{semilogyaxis}
    \end{tikzpicture}
    
    \caption{JTD decoding error probability versus target DOA \(\omega_s\) for $M=13$ antenna ULA and BCA arrays with $\omega_c = 0$ and at \(\mathrm{SNR} = 10\) dB.}
    \label{fig:jtd-delta-sweep-decoding}
    \vspace{-15pt}
\end{figure}
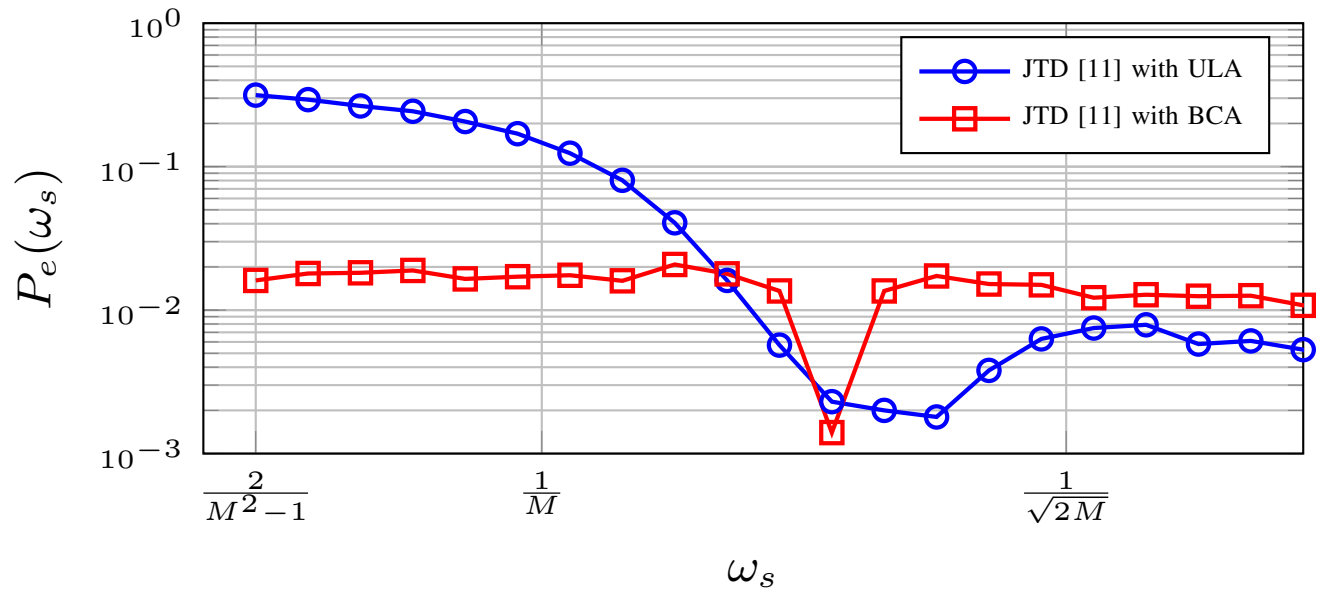

In \Cref{fig:jtd-delta-sweep-decoding}, we further examine the close-separation behavior predicted by the DTS analysis using the JTD receiver. In this experiment, the UE direction is fixed at \(\omega_c=0\), and the target direction is chosen according to $\omega_s = \Delta,$
where $\Delta = \frac{2k}{M^2-1},\ k=1,\ldots,25$ and the SNR is fixed at \(10\) dB. In this figure, we can observe that BCA is significantly better than ULA in the closely-separated regime between $\frac{2}{M^2-1}$ to $\frac{1}{M}$. While the performance of ULA fluctuates in the intermediate region of $\frac{1}{M}\leq \Delta \leq \frac{1}{\sqrt{2M}}$, BCA maintains low and nearly constant decoding error probability for all $\Delta$ considered. In the well-separated regime of $\Delta>\frac{1}{\sqrt{2M}}$ we see that performance of ULA stays close to BCA. 

\vspace{-10pt}
\section{Conclusion}
\vspace{-4pt}
\label{Sec_Conclusion}
We studied bistatic ISAC through a subspace-coding lens, showing that array geometry governs the coupling between communication and sensing via the subspace distance between their steering-vector subspaces. We showed that larger subspace distance improves data decoding reliability by mitigating sensing-induced interference. Specifically, sparse geometries such as BCAs—interpretable as spatial subspace codes—offer favorable ISAC performance, especially when the DOA difference between the sensing and communications channels is small. This suggests subspace coding as a unifying principle for ISAC array design.

\vspace{-0.21cm}
\bibliography{references}
\bibliographystyle{ieeetr}

\appendices\crefalias{section}{appendix}
\renewcommand{\thesectiondis}[2]{\Alph{section}:}

\end{document}